\documentclass[12pt]{article}
\usepackage[latin1]{inputenc}
\usepackage{pstricks}
\usepackage{amsmath}
\usepackage{float}
\usepackage{color}
\input epsf
\usepackage{amssymb}
\usepackage[dvips]{graphicx,psfrag}
\newcommand{\be}{\begin{equation}}
\newcommand{\ee}{\end{equation}}
\newcommand{\bea}{\begin{eqnarray}}
\newcommand{\eea}{\end{eqnarray}}
\newcommand{\lb}{\label}
\begin{document}
\begin{titlepage}
%\title{On some universes with peculiar expansion}
\title{Past and future of some universes}
\author{ D. Polarski\thanks{email:david.polarski@univ-montp2.fr}\\
\hfill\\
Universit\'e Montpellier 2 \& CNRS, Laboratoire Charles Coulomb\\ 
UMR 5221, F-34095 Montpellier,France }

\pagestyle{plain}
\date{\today}

\maketitle

\begin{abstract}
We consider a class of toy models where a spatially flat universe is filled with 
a perfect fluid. The dynamics is found exactly for all these models. In one family, 
the perfect fluid is of the phantom type and we find that the universe is first 
contracting and then expanding while the dynamics is always accelerated. 
In a second family, the universe is first in an accelerated expansion stage, then 
in a decelerated expansion stage until it reaches a turning point after which it 
contracts in a decelerated way (increasing contraction rate) followed by another 
accelerated stage (decreasing contraction rate). We also consider the possibility 
to embed this perfect fluid in a realistic cosmology. The first family cannot be 
viable in a conventional big bang universe and requires a rebound in the very early 
universe. The second family is viable in the range $0<1+w_{DE,0}\lesssim 0.09$ 
for a spatially closed universe with a curvature satisfying current bounds.  
Though many of the models in this family cannot be distinguished today from a universe 
dominated by a cosmological constant, the present accelerated expansion is transient 
and these universes will reach a turning point in the future before entering a 
contraction phase.   
\end{abstract}

\end{titlepage}

%%%%%%%%%%%%%%%%%%%%%%%%%%%%%%%%%%%%%%%%%%%%%%%%%%%%%%%%%%%%%%%%%%%%%%%%%%%%%%%%
%%%%%%%%%%%%%%%%%%%%%%%%%%%%%%%%%%%%%%%%%%%%%%%%%%%%%%%%%%%%%%%%%%%%%%%%%%%%%%%%

\section{Introduction}

Since the discovery of the universe present accelerated expansion rate \cite{P97}, 
our view of expanding universes has been durably affected. There has been a lot of interest 
in models producing the accelerated expansion \cite{SS00} as we see it at the present 
time. The simplest models introduce dark energy (DE) in the form of a comoving 
perfect fluid with a sufficiently negative pressure today. A very simple and appealing 
example is provided by a cosmological constant $\Lambda$ which agrees well with existing 
data. Remarkably, this corresponds to a constant energy density and a constant pressure 
satisfying $\rho_{\Lambda}+p_{\Lambda}=0$ or $w_{\Lambda}=-1$. Leaving aside the conceptual 
problems raised by the existence of such a tiny cosmological constant, it remains to be seen 
whether more accurate observations will force us into a paradigm shift. In any case it remains 
an important issue to look for other viable alternatives where the energy density and pressure 
will have a different physical interpretation and behaviour. In particular it may well 
be that no DE component is needed at all if the present accelerated expansion is caused by 
modifications of gravity on cosmic scales.    
A miscroscopic interpretation of the DE energy density and pressure is given in models 
like quintessence \cite{RPW88}, a time-dependent homogeneous scalar field minimally 
coupled to gravity. Through and Einsteinian interpretation of the modified Friedmann equations, 
a microscopic description of an effective DE energy density and pressure can be given as well 
for some models going beyond General Relativity, (see e.g.\cite{GMP08} in the context of $f(R)$ 
models) where gravity is modified on cosmic scales.
This radical departure from conventional expansion histories has also brought interest 
for DE of the phantom type ($w_{DE}<-1$) \cite{C02} and for the future of our 
universe \cite{S99}. As we will see below, our work touch upon both issues.

Here we introduce a class of phenomenological perfect fluids and we study the 
corresponding universe dynamics. The crucial property is that the energy density 
of this perfect fluid can vanish in the course of expansion or contraction, depending on 
the specific model. We consider toy models where spatially flat universes contain only 
this perfect fluid and we solve for their dynamics. As expected this gives rise to very 
peculiar dynamics.
In one family the perfect fluid is forever of the phantom type and it will tend to a 
cosmological constant in the asymptotic future after passing through a bounce. 
In the second family the universe expands till a turning point after which it starts 
contracting.  
We also consider the possibility to include such a component in a realistic 
cosmology. We will find that viable universes can exist with a radical 
departure from standard cosmological scenarios either in the past, for the perfect 
fluid of the phantom type, or in the future for the second family. The departure 
in the universe past in the first case comes from a conceptual problem and requires 
probably to go beyond General Relativity in the early universe. However for the second family 
a universe with a peculiar future can be obtained in the framework of General Relativity.

%%%%%%%%%%%%%%%%%%%%%%%%%%%%%%%%%%%%%%%%%%%%%%%%%%%%%%%%%%%%%%%%%%%%%%%%%%%%%%%%%%%%%%%%%%%
%%%%%%%%%%%%%%%%%%%%%%%%%%%%%%%%%%%%%%%%%%%%%%%%%%%%%%%%%%%%%%%%%%%%%%%%%%%%%%%%%%%%%%%%%%%
\section{The model}

We consider a Friedmann-Lema\^\i tre- Robertson-Walker (FLRW) spatially flat 
universe filled with a comoving pefect fluid of energy density $\rho$ and 
pressure $p$. We have the following equations 
\bea
H^2 &=& \frac{8\pi G}{3}~\rho \lb{FR1}\\
{\dot H} &=& -4\pi G ~(\rho + p)~, \lb{FR2}
\eea
The time evolution of the perfect fluid 
\be
{\dot \rho} = -3 H (\rho + p) = -3 H (1 + w)\rho~,\lb{drho}
\ee
is contained in (\ref{FR1}),(\ref{FR2}) and we have introduced the equation of 
state (EoS) parameter $w\equiv p/\rho$. 
Let us start with a perfect fluid satisfying 
\be
p = {\rm constant}~.\lb{p1}
\ee
It is easy to derive from \eqref{FR1},\eqref{FR2} 
\be
d(\rho~a^3) = -p ~da^3 = -d(p~a^3) ~,
\ee
which leads to 
\be
\rho~a^3 = -p~a^3 - B~, 
\ee
or 
\be
\rho = -p - \frac{B}{a^3}\ge 0~. 
\ee
Setting $p=-A$, we can write 
\be
\rho = A - \frac{B}{a^3}~. \lb{rho1}
\ee 
It is easy to understand why both terms can appear in \eqref{rho1}: from 
\eqref{drho}, the first term leads to $\rho$ and $p$ being constant while 
the second term corresponds to $p=0$. The important point here is that $B$ 
is an arbitrary integration constant. It can vanish and while it does not, 
it can have \emph{arbitrary} sign, the sign  is not specified by \eqref{p1}. 

The case $B>0$ corresponds to a phantom equation of state, $w<-1$. This will 
always be the case, there is no crossing of the phantom divide line. 
We have in particular from \eqref{FR2}
\be
{\dot H} \ge 0~.\lb{dH}   
\ee
It is not possible in this case to interpret \eqref{rho1} as a cosmological plus 
some dust because the additional dustlike component would have a negative energy 
density due to the ``wrong'' sign of $B$. It is the sum of both terms in \eqref{rho1} 
that gives a physically sound energy density satisfying $\rho \ge 0$.  

Solving for the scale factor $a$ as a function of $t$ we obtain
\be
a = a_B \cosh^{\frac{2}{3}}\left[ \frac{3}{2}~H_{\infty}(t-t_B)\right]~.\lb{a1}
\ee
Here $a_B$ is the minimal value of the scale factor reached at time $t_B$ where the 
Hubble parameter $H$ vanishes. Equation \eqref{rho1} can be recast as 
\be
\rho = A~\left[ 1 - \left( \frac{a_B}{a}\right)^3 \right]~. \lb{rho1B}
\ee
Though the universe we consider is spatially flat, 
we get a bouncing universe with contraction at times $t<t_B$ and expansion at 
times $t>t_B$. We have defined
\be
H^2_{\infty} \equiv \frac{8 \pi G}{3} A~,
\ee
which represents the value of $H$ in the asymptotic future. The Hubble parameter 
$H$ in function of $t$ is readily obtained from \eqref{a1}, viz.
\be
H = H_{\infty} \tanh \left[ \frac{3}{2}~H_{\infty}(t-t_B)\right]~,\lb{H1}
\ee
from which we see again that $H$ tends to $H_{\infty}$ for $t\to \infty$.
To summarize, our universe is always in a phantom stage, it is first contracting 
then expanding and tends to a pure de Sitter universe with $H= H_{\infty}$ for 
$t\to \infty$.

It is interesting to compare \eqref{a1} and \eqref{H1} with the solutions obtained 
in a matter dominated universe with a cosmological constant $\Lambda$
\bea
a &=& \left( \frac{\Omega_{m,0}}{\Omega_{\Lambda,0}} \right)^{\frac{1}{3}} a_0 
                           \sinh^{\frac{2}{3}} \left( \frac{3}{2}~H_{\infty} t \right)\\
H &=& H_{\infty} \coth \left( \frac{3}{2}~H_{\infty} t \right)~,
\eea
with $H^2_{\infty}\equiv \frac{\Lambda}{3}$. In this case we have of course a 
``Big Bang'' singularity with $H\to \infty$ as $t\to 0$. It is the past which is 
radically different for both universes.

We can look for more general behaviours 
\be
\rho = A - \frac{B}{a^n}~~~~~~~~~~~~~~~~~~~~~~~~~A,B>0~.\lb{rhon}
\ee
We have in particular $A - \rho>0$.
It is easy to obtain the corresponding pressure $p$
\be
p = - \rho - \frac{n}{3}~(A - \rho)~. 
\ee
We consider first the case $n>0$.
We check that a phantom type equation of state is obtained for $n>0$, a fact 
already clear by inspection of \eqref{rhon}. Hence a universe filled 
with a perfect fluid that evolves according to \eqref{rhon} undergoes only 
accelerated expansion. 
We can generalize the expressions \eqref{a1}, \eqref{H1}, viz.
\bea
a &=& a_B \cosh^{\frac{2}{n}}\left[ \frac{n}{2}~H_{\infty}(t-t_B)\right]~.\lb{an}\\
H &=& H_{\infty} \tanh \left[ \frac{n}{2}~H_{\infty}(t-t_B)\right]~.\lb{Hn}
\eea
Actually these expressions apply to negative $n$ as well. 
It is interesting to consider also the case $n<0$ as well. Of course, now we have no phantom type 
equation of state and $\dot{H}<0$ The universe is a de Sitter space in the infinite 
past and we can call its (constant) Hubble parameter $H_{\infty}$. It will reach some 
turning point in the future with expansion followed by a contraction stage. We have 
now (for $n<0$)
\bea
a &=& a_B \cosh^{-\frac{2}{|n|}}\left[ \frac{|n|}{2}~H_{\infty}(t-t_B)\right]~.\lb{anneg}\\
H &=& H_{\infty} \tanh \left[ -\frac{|n|}{2}~H_{\infty}(t-t_B)\right]~.\lb{Hnneg}
\eea
It is seen that $a_{-n}=a^{-1}_{n}$ from which we obtain, in accordance with eqs. \eqref{an} 
and \eqref{anneg}
\be
H_{-n} = - H_{n}~.
\ee
For $n<0$, $\dot{H}<0$ so that the expansion is not always accelerated. Actually the 
evolution is now very different. The universe is first expanding and then it starts contracting 
at time $t_B$. In the asymptotic past, $t\to -\infty$, the universe tends to an expanding de 
Sitter space while it tends to a contracting de Sitter space in the asymptotic future, 
$t\to \infty$. During some (symmetric) interval around $t_B$ one has $\ddot{a}<0$ similarly 
to a spatially closed universe going over from expansion to contraction.   
At some time $t_a<t_B$, the evolution goes over from accelerated expansion 
to decelerated expansion on the interval $t_a = t_B-\Delta t < t < t_B$ with 
$\Delta t\equiv t_B - t_a$. 
This is followed by decelerated contraction on the interval $t_B < t < t_B + \Delta t$. For 
$t > t_B + \Delta t$ the universe undergoes accelerated contraction.  
The interval $\Delta t$ is easily found, viz.   
\be
\sinh^2 \left[\frac{|n|}{2}~H_{\infty}~\Delta t \right] = \frac{|n|}{2}~. \lb{Delta}
\ee 
It is seen that the interval $\Delta t$ depends on $|n|$. 
Hence for $n<0$, the universe evolution can be suumarized as follows:
\bea 
\ddot{a} & > & 0~~~~~~~~~~~~~~~~~~~~~~|t - t_B| > \Delta t \\
\ddot{a} & < & 0~~~~~~~~~~~~~~~~~~~~~~|t - t_B| < \Delta t~,
\eea 
with $\Delta t$ found from \eqref{Delta}, and further 
\bea
\dot{a} & > & 0~~~~~~~~~~~~~~~~~~~~~~~t < t_B \\
\dot{a} & < & 0~~~~~~~~~~~~~~~~~~~~~~~t > t_B~.
\eea 
Let us turn our attention to the equation of state (EoS) parameter $w=\frac{p}{\rho}$. 
It is straightforward to derive 
\be
w = - 1 - \frac{n}{3} \left( \frac{A}{\rho} - 1 \right)~,
\ee
which shows that $w$ diverges whenever $H=0$. This has to be the case as we show now. 

Assuming a contracting universe ($n>0$) starts from some regular initial 
conditions at time $t_1$ and reaches a bounce at time $t_B$, the following equality 
must hold from (\ref{FR1}),(\ref{FR2})
\be
\int_{t_1}^{t_B} dt~(1 + w) = - \infty~.\lb{cond1}
\ee
Clearly, this equality cannot be satisfied for $t_B<\infty$ if $w$ is regular. 
Assuming an expanding universe with $n<0$ starts from some regular initial 
conditions at time $t_1$ and reaches a turning point at time $t_B$, we get 
analogously
\be
\int_{t_1}^{t_B} dt~(1 + w) = \infty~,\lb{cond1b}
\ee
leading to a similar conclusion. 

\section{Embedding in a realistic cosmology}

A further interesting issue concerns the embedding of our perfect fluid 
in a realistic cosmology. This is what will be investigated now. 
It is natural to assign to our perfect fluid the role of dark energy 
responsible for the late-time accelerated expansion of the universe. Hence 
our fluid will be called below dark energy with energy density $\rho_{DE}$ 
and pressure $p_{DE}$.
For our discussion it is convenient to introduce the notation
\be
B = B_0~a_0^n~,
\ee
so that \eqref{rhon} can be recast in the form
\be
\rho = A - B_0~\left( \frac{a}{a_0} \right)^{-n}~, \lb{rhoB0}
\ee
where $B_0$ has dimensions of energy with $B_0=A-\rho_0$. It is easy to show 
\be
\frac{a_B}{a_0} = \left( \frac{B_0}{A} \right)^{\frac{1}{n}}~. 
\ee
When $n>0$, $a_B$ corresponds to a bounce with $a_B\to 0$ for $B_0\to 0$. When 
$n<0$, $a_B$ corresponds to a turning point with $a_B\to \infty$ for $B_0\to 0$.
The Friedmann equations yield now
\bea
H^2 &=& \frac{8\pi G}{3} ~\left[ \rho_m + \rho_r + A - B_0 \left( \frac{a}{a_0} \right)^{-n}\right] 
                  + \Omega_{k,0}~H^2_0~\frac{a^2_0}{a^2}             \lb{FR1b}\\
\frac{\ddot a}{a} &=& -\frac{4\pi G}{3}\left[ \rho_m + 3 \rho_r -2\rho_{DE} -n (A-\rho_{DE}) \right]\lb{FR2b}
\eea
where we have allowed for spatially non-flat universes. 
We finally note that $n=0$ corresponds to a cosmological constant, only 
the difference $A-B_0$ matters with $8\pi G~(A-B_0)\equiv \Lambda$ and this case 
needs no further consideration.

\subsection{$n>0$}

When $n>0$, the universe will tend in the future to a de Sitter space. The dark energy 
sector is of the phantom type, it tends asymptotically to a cosmological constant from 
below ($w_{DE}<-1$) which is not possible for usual quintessence (a minimally coupled scalar 
field). We have curiously a ``growing'' rather than ``decaying'' effective cosmological 
constant.  

The past of the universe must be changed compared to the standard big bang scenario. 
While our perfect fluid has no influence of the universe dynamics in the past -- actually it 
becommes dynamically significant only at $z<1$ -- the universe is not allowed to cross in 
the past the value $a=a_B$ if we insist on the condition $\rho_{DE}\ge 0$. 
On the other, if we want a universe where the main features of the hot big bang universe 
is retained, $B_0$ has to be small enough so that 
\be
\left( \frac{B_0}{A} \right)^{\frac{1}{n}} \ll 10^{-10}~, \lb{ineq1}           
\ee  
which requires a fine tuning of the model in addition to the fine tuning of 
$A$ (the cosmic coincidence). Viability of such a universe requires further that the 
universe has a bounce at $a>a_B$ when $\rho_{DE}$ is still positive. This requires to go 
beyond General Relativity in the very early universe (e.g. \cite{Ve03}. In any case, due 
to \eqref{ineq1}, such a universe is undistinguishable observationally from a universe 
with a cosmological constant. 
We believe the main merit of this model is that it provides an interesting conceptual 
problem.  

\subsection{$n<0$}

When $n<0$, the primordial universe can be governed by equations \eqref{FR1b},\eqref{FR2b} 
with a standard big bang cosmology. The universe future is changed with a turning point 
after which the universe will be constracting. 

The danger with such a universe is that $a_B$ could lie in our past which is forbidden 
as further expansion would force $\rho_{DE}$ to be negative and the universe could not
reach the present day expansion stage.  
The condition $B_0<A$ is obviously sufficient to guarantee that $a_B$ lies in the future, 
$a_B>a_0$. 
Taking $\frac{B_0}{A}\ll 1$ will push $a_B$ (far) in the future. In such a model dark energy 
is represented by a slowly ``decaying'' cosmological constant. Actually, even if $a_B\gg a_0$, 
once the point $a=a_B$ is reached, further expansion -- which is unavoidable because of the 
presence of dustlike matter (and radiation) however tiny their energy densities -- would 
inevitably imply $\rho_{DE}<0$. This is reminiscent of the situation encountered with a 
Big-Brake singularity in the future \cite{KGKGP10} if one is willing to include dustlike matter. 
 
The only way to avoid this is to have a closed universe. Let us call $a_1$ the value of the scale 
factor at which the dustlike term and the curvature term cancel each other in the right hand 
side of the Friedmann equation. We then have 
\be
\frac{a_1}{a_0} = \frac{\Omega_{m,0}}{|\Omega_{k,0}|}~. 
\ee
Inclusion of radiation is possible but it will have negligible influence. 
To ensure that the turning point is reached before $\rho_{DE}$ becomes negative, we must 
impose $a_1\le a_B$. This can be expressed as a lower bound on $|\Omega_{k,0}|$ in 
function of the model perameters  
\be
|\Omega_{k,0}|\ge \Omega_{m,0}~\left( \frac{B_0}{A} \right)^{\frac{1}{|n|}}~.  \lb{ineq2}
\ee
In a realistic cosmology, $-0.0065 \le \Omega_{k,0} \le 0.0012$ (95\% C.L.) with central 
value $\Omega_{k,0}=-0.0027$ \cite{WMAP12}. So there is preference for a slightly 
spatially closed universe which is exactly what we need. Condition \eqref{ineq2} 
can obviously be satisfied for a large range of parameter values satisfying 
\be
\frac{\Omega_{B,0}}{\Omega_{A,0}}\lesssim  \left( 0.026 \right)^{|n|}  ~. \lb{ineq2a}
\ee    
where we have defined $\Omega_{B,0}\equiv \frac{B_0}{\rho_{cr,0}}$ and 
$\Omega_{A,0}\equiv \frac{A}{\rho_{cr,0}}$ and we have taken $\Omega_{m,0}\simeq 0.25$. 
Violation of \eqref{ineq2} for some set of parameter values implies that the 
observational upper bound on spatial curvature is not large enough to ensure 
$\rho_{DE}\ge 0$ in the future.  

Inequality \eqref{ineq2a} can be recast into 
\be
\frac{\Omega_{B,0}}{\Omega_{DE,0}}\lesssim \frac{1}{ (0.026)^{-|n|} - 1 }  \lb{ineq2b}
\ee
For example for $|n|=2$ we have $\Omega_{B,0}\lesssim 0.00067~\Omega_{DE,0}$, for 
$|n|=0.5$, one obtains $\Omega_{B,0}\lesssim 0.192~\Omega_{DE,0}$. For $|n|\lesssim 0.5$, 
$\Omega_{B,0}$ can be substantial while it becomes negiligible for $|n|\gtrsim 2$. 

On the other hand, the expansion must be accelerated at the present epoch which yields
\be
\Omega_{B,0} < \frac{2 \Omega_{DE,0} - \Omega_{m,0}}{|n|} \lesssim \frac{1.25}{|n|}~,\lb{ineq3}
\ee
Condition \eqref{ineq3} corresponds to $w_{DE,0}< -\frac{1}{3}\Omega^{-1}_{DE,0}$ (neglecting 
here $\Omega_{k,0}$). We know however from the observations that $w_{DE,0}$ is significantly 
smaller. Hence a much stronger condition comes from the smallness of $1+w_{DE,0}$. Indeed we 
have 
\be
\frac{\Omega_{B,0}}{\Omega_{DE,0}} = \frac{3}{|n|}~(1 + w_{DE,0})~. \lb{ineq4}
\ee
If we take $1 + w_{DE,0}\lesssim 0.1$ and $\Omega_{DE,0}\simeq 0.75$, we obtain 
$\Omega_{B,0}\lesssim \frac{0.225}{|n|}$. As expected, this is much stronger than 
\eqref{ineq3}.  

Combining condition \eqref{ineq2b} with \eqref{ineq4}, we get a new inequality
\be
1 + w_{DE,0}\lesssim \frac{|n|}{3}~\frac{1}{ (0.026)^{-|n|} - 1 }\lb{ineq5}
\ee
It is interesting that \eqref{ineq5} implies 
\be
1 + w_{DE,0}\lesssim 0.09~.
\ee
Hence this model is able to cover the range of values relevant for $1 + w_{DE,0}>0$. 
As the observational bound on $|\Omega_{k,0}|$ will be more stringent, the range of 
allowed values for  $1 + w_{DE,0}>0$ will decrease. 

Let us consider the regime $|n|\ge 2$. In that case we have 
$\Omega_{B,0}<6\times 10^{-4}~\Omega_{A,0}$, 
$\Omega_{B,0}<6\times 10^{-4}~\Omega_{DE,0}$ and $\Omega_{A,0}\simeq \Omega_{DE,0}$. 
We also get in this case from \eqref{ineq5} that $1 + w_{DE,0}< 4\times 10^{-4}$. 
These models cannot be distinguished from a cosmological constant model with 
$\Omega_{A,0}\simeq \Omega_{\Lambda,0}$.  
Actually for models with $|n|\gtrsim 0.95$, the quantity $1 + w_{DE,0}$ is strongly 
constrained with $1 + w_{DE,0}\lesssim 0.01$. For large $|n|$ values both quantities 
$\frac{\Omega_{B,0}}{\Omega_{DE,0}}$ and $\frac{\Omega_{B,0}}{\Omega_{A,0}}$ are 
much smaller than one. 
It is possible to have substantially higher values for $1 + w_{DE,0}$ when $|n|\to 0$. 
For $|n|\approx 0.5$, the quantity $1 + w_{DE,0}$ is allowed to be inside the range   
$0<1 + w_{DE,0}\lesssim 0.03$. For lower nonzero values $|n|\approx 0$, 
$1 + w_{DE,0}$ can be as large as $0.09$. 
In this range of small $|n|$ values, $\frac{\Omega_{B,0}}{\Omega_{DE,0}}$ is allowed to be 
much larger than $1$ with $\Omega_{B,0}\simeq \Omega_{A,0}$ (and $\Omega_{B,0}<\Omega_{A,0}, 
~\Omega_{A,0} - \Omega_{B,0}= \Omega_{DE,0}\approx 0.75$). The quantity $\Omega_{B,0}$ can 
be large because $|n|$ is very small, and therefore the time evolution of $\rho_{DE}$ is 
slow enough so that one can still obtain $a_B\gg a_0$ and $a_B\ge a_1$. 

If observations show evidence for $w_{DE,0}$ significantly higher than $-1$ but still 
satisfying $w_{DE,0}\lesssim 0.09$, then only the low $|n|$ values remain viable. 
In all these universes the late-time accelerated expansion is a 
transient phenomenon. It is followed by decelerated expansion and eventual contraction.

\section{Summary}

We have considered spatially flat toy models universes filled with a particular 
perfect fluid and we have solved exactly the corresponding dynamics. In one 
family of models ($n>0$) the perfect fluid is of the phantom type and an ever 
accelerating dynamics is obtained: the universe is first contracting, then passing 
through a bounce after which it is expanding. We have exponential contraction 
in the infinite past and the universe tends to a de Sitter expanding universe in the 
asymptotic future. In the second family of models ($n<0$) the universe is exponentially 
expanding in the asymptotic past and exponentially contracting in the asymptotic future. 
At some time ($t_a$) the expansion becomes decelerated, the scale factor then reaches a 
maximum followed by a contracting phase, first decelerating then accelerating (decreasing 
contraction rate). 
We have also considered the possibily to embed our perfect fluid in a realistic cosmology 
in which case it can play the role of dark energy. 
When ($n>0$) we have the strange situation that while the corresponding energy density is 
negligible already at low redshifts, a standard big bang singularity would force the energy 
density of our perfect fluid to become negative. We conjecture that this could be avoided 
by having a rebound before this happens, which could be the case if we go beyond General 
Relativity in the very early universe. On the other hand, at the present time this universe 
cannot be distinguished from a $\Lambda$ dominated universe and it tends to a de Sitter 
(expanding) universe in the future with $1+w_{DE}<0$.  

As for the other family of solutions ($n<0$), a slightly spatially closed universe is 
required in order to avoid $\rho_{DE}$ to become negative in our future. Interestingly, 
using the current observational bound on $|\Omega_{k,0}|$ we find that all universes 
with $0<1 + w_{DE,0}\lesssim 0.09$ can be accomodated by this model. The maximal values 
$1 + w_{DE,0}\simeq 0.09$ are obtained for small values $0\ne |n|\ll 1$. 
For $|n|\ge 2$, we have essentially a $\Lambda$ dominated universe with 
$\Omega_{B,0}\ll \Omega_{A,0}\simeq \Omega_{\Lambda,0}$ and $1 + w_{DE,0}\ll 1$. 
The strong inequality $\Omega_{B,0}\ll \Omega_{A,0}$ can be seen as an additional fine 
tuning. These models certainly do not pretend to get rid of any of the conceptual problems 
of the $\Lambda$ model, what is interesting is that they can provide a radically different 
future. Because the effective cosmological constant is slowly decaying,
the accelerated expansion is a transient phenomenon and the present expansion 
stage will be followed by a contracting phase even though this universe cannot be distinguished 
today from a $\Lambda$ dominated universe. The future of our universe may depend crucially 
on the precise form of the dark energy sector even if it cannot be distinguished today from 
a cosmological constant, and our model provides just such a example. 

%%%%%%%%%%%%%%%%%%%%%%%%%%%%%%%%
%\section*{Acknowledgments}

%%%%%%%%%%%%%%%%%%%%%%%%%%%%%%%%
  
%%%%%%%%%%%%%%%%%%%%%%%%%%%%%%%%%%%%%%%%%%%%%%%%%%%%%%%%%%%%%%%%%%%%%%%%%%%%%%%%%%%%%%%%%%%%%%
%%%%%%%%%%%%%%%%%%%%%%%%%%%%%%%%%%%%%%%%%%%%%%%%%%%%%%%%%%%%%%%%%%%%%%%%%%%%%%%%%%%%%%%%%%%%%%


\begin{thebibliography}{99}

\bibitem{P97} 
 S.J. Perlmutter et al.,\emph{Ap. J.} {\bf 483} 565 (1997), \emph{Nature} \textbf{391}, 1998, 51;
 A. G.~Riess, A. V.~Filippenko, P.~Challis {\it et al.},\emph{ Astron.~J.} \textbf{116}, 1998, 1009;
 P.~Astier, J.~Guy, N.~Regnault {\it et al.}, \emph{ Astron. Astroph.} \textbf{447}, 2006, 31



\bibitem{SS00} 
 V. Sahni, A. A. Starobinsky, Int. J. Mod. Phys. D{\bf 9}, 373 (2000);   
 T. Padmanabhan, Phys. Rep.{\bf 380}, 235 (2003);
 E. J. Copeland, M. Sami and S. Tsujikawa, Int. J. Mod. Phys. D{\bf 15}, 1753 (2006);
 V. Sahni, A. A. Starobinsky, Int. J. Mod. Phys. D{\bf 15}, 2105 (2006);
 M. Li, X.-D. Li, S. Wang, Y. Wang, Commun. Theor. Phys. {\bf 56} 525 (2011) [arXiv:1103.5870];
 D. Weinberg, M. Mortonson, D. Eisenstein, C. Hirata, A. Riess, E. Rozo [arXiv:1201.2434];
 L. Amendola, S. Tsujikawa, \emph{Dark Energy: Theory and Observations}, Cambridge University Press, 2010.

\bibitem{RPW88}
 B. Ratra and P.J.E. Peebles, Phys. Rev. D\textbf{37}, 3406 (1988);
 C. Wetterich, Nucl. Phys. B\textbf{302}, 668 (1988); 
 P. G. Ferreira, M. Joyce, Phys.Rev. D\textbf{58}, 023503 (1998);
 A. Liddle and R. Scherrer, Phys. Rev. D\textbf{59}, 023509 (1999);
 I. Zlatev, L. Wang and P.J. Steinhardt, Phys. Rev. Lett.\textbf{82}, 896 (1999).

\bibitem{GMP08}
 R. Gannouji, B. Moraes, D. Polarski, JCAP \textbf{0902}, 034 (2009).

\bibitem{C02} 
 R. Caldwell, Phys. Lett. B \textbf{545}, 23 (2002);
 M. P. Dabrowski, T. Stachowiak, M. Szydlowski, Phys. Rev. D \textbf{68}, 103519 (2003);
 B. Boisseau, G. Esposito-Far\`ese, D. Polarski and A.A. Starobinsky, Phys. Rev. Lett.\textbf{85}, 2236 (2000);
 R. Gannouji, D. Polarski, A. Ranquet, A. A. Starobinsky, JCAP \textbf{0609}, 016 (2006);
 M.P. Dabrowski, C. Kiefer, B. Sandhofer, Phys.Rev. D \textbf{74}, 044022 (2006).

\bibitem{S99}
 A. A. Starobinsky, Grav.Cosmol.\textbf{6}, 157 (2000);
 R. R. Caldwell, M. Kamionkowski, N. N. Weinberg, Phys.Rev.Lett.\textbf{91}, 071301 (2003);
 S. Nojiri, S. D. Odintsov, S. Tsujikawa, Phys.Rev. D\textbf{71}, 063004 (2005)
 P. H. Frampton, K. J. Ludwick, R. J. Scherrer, Phys.Rev. D\textbf{84}, 063003 (2011)
 
\bibitem{Ve03}
 M. Gasperini, G. Veneziano, Phys.Rept.\textbf{373}, 1 (2003) 
 J. Martin, P. Peter, Phys.Rev. D\textbf{68}, 103517 (2003)
 J. Grain, A. Barrau, T. Cailleteau, J. Mielczarek, Phys.Rev. D\textbf{82}, 123520 (2010)

\bibitem{KGKGP10}	
Z. Keresztes, L. A. Gergely, A. Yu. Kamenshchik, V. Gorini, D. Polarski, Phys.Rev. D\textbf{82} 123534 (2010)

\bibitem{WMAP12}
 G. Hinshaw, D. Larson, E. Komatsu, D. N. Spergel, C. L. Bennett, J. Dunkley, M. R. Nolta, M. Halpern, 
 R. S. Hill, N. Odegard, L. Page, K. M. Smith, J. L. Weiland, B. Gold, N. Jarosik, A. Kogut, M. Limon, 
 S. S. Meyer, G. S. Tucker, E. Wollack, E. L. Wright, arXiv:1212.5226

\end{thebibliography}
\end{document}